\documentclass[12pt]{iopart}

\usepackage{graphicx}

\begin{document}

\title[Axial light emission in a parallel plate dc micro discharge]{Axial light emission and Ar metastable densities in a parallel plate dc micro discharge in steady state and transient regimes}

\author{T Kuschel$^1$, B Niermann$^1$, I Stefanovi$\acute{\textrm{c}}^1$, M B\"oke$^1$, N $\check{\textrm{S}}$koro$^2$, D Mari$\acute{\textrm{c}}^2$, Z Lj Petrovi$\acute{\textrm{c}}^2$ and J Winter$^1$}

\address{$^1$Ruhr-Universit\"at Bochum, Institute for Experimental Physics II, Universit\"atsstra\ss e 150, 44780 Bochum, Germany}
\address{$^2$Institute of Physics, University of Belgrade, POB 68, 11080 Belgrade, Serbia}

\ead{Thomas.Kuschel@rub.de}

\begin{abstract}
Axial emission profiles in a parallel plate dc micro discharge (feedgas: argon; discharge gap $d=1\,$mm; pressure $p = 10\,$Torr) were studied by means of time resolved imaging with a fast ICCD camera. Additionally, volt-ampere ($V$-$A$) characteristics were recorded and Ar* metastable densities were measured by tunable diode laser absorption spectroscopy (TDLAS). Axial emission profiles in the steady state regime are similar to corresponding profiles in standard size discharges ($d\approx 1\,$cm, $p \approx 1\,$Torr). For some discharge conditions relaxation oscillations are present when the micro discharge switches periodically between the low current Townsend-like mode and the normal glow. At the same time the axial emission profile shows transient behavior, starting with peak distribution at the anode, which gradually moves towards the cathode during the normal glow. The development of argon metastable densities highly correlates with the oscillating discharge current. Gas temperatures in the low current Townsend-like mode ($T_{\textrm{\scriptsize{g}}} = 320$-$400\,$K) and the high current glow mode ($T_{\textrm{\scriptsize{g}}} = 469$-$526\,$K) were determined by the broadening of the recorded spectral profiles as a function of the discharge current.           
\end{abstract}


\submitto{\PSST}

\maketitle

\section{Introduction}

Microplasmas have recently become a focus of research due to the wide range of their possible applications \cite{Becker}. Different kinds of microplasma sources have been proposed: micro atmospheric pressure plasma jet ($\mu$-APPJ) \cite{Gathen}, micro hollow cathode \cite{Schoenbach}, and large arrays of micro discharges with dielectric barrier \cite{Eden} are some of the most popular. However, very few studies of parallel plate micro discharges exist at all \cite{Petrovic2,Wang,Greenan}. Due to their simple geometry the parallel plate micro discharges can be used as an ideal benchmark for different plasma models and for testing the similarities between large scale low pressure discharges and micro discharges.

Reproducible and stable discharge conditions are of high importance to realize reliable applications. However, these conditions are usually not achievable over the full operation range. Observations of self-pulsing regimes were amongst others reported in micro hollow cathode discharges \cite{Aubert}, micro thin-cathode discharges \cite{Du}, micro plasma jets \cite{Walsh} and recently in parallel plate micro discharges \cite{Stefanovic2}. Numerous experiments in standard size parallel plate dc discharges ($d\approx 1\,$cm, $p \approx 1\,$Torr) have shown that different instabilities may occur \cite{Petrovic1,Jelenkovic,Phelps,Stefanovic} and the discharge does not operate in a stable regime but moves through a transient phase, switching repetitively from low to high current mode. With the development of ICCD cameras time resolved measurements of discharge transients became possible \cite{Maric1,Maric3}. In our previous work we have shown first 2D time integrated recordings of the axial light emission in a parallel plate dc micro discharge \cite{Kuschel}.

In this contribution we continue our studies and show time resolved 2D recordings of a parallel plate dc micro discharge ($d=1\,$mm, $p = 10\,$Torr) during relaxation oscillations. ICCD camera images are correlated with current and voltage measurements to gain a better understanding of the formation of space charge effects and the cathode fall formation. Axial light distributions under steady state conditions (static $V$-$A$ characteristics) have been measured and used to compare with discharge transients.

In addition we have applied tunable diode laser absorption spectroscopy (TDLAS) to record the spectral profiles of the lowest argon metastable state, deducing Ar* (1s$_5$) densities for various discharge conditions. Due to their long lifetime, atoms in metastable states are a reservoir of energy in the discharge, and stepwise ionization through these states is known to be an important ionization mechanism in rare gas plasmas, especially when the electron temperature is low. This is especially important in micro discharges as due to $jd^2$ scaling current densities as well as metastable densities may be quite high even in the low current Townsend regime, therewith leading to three body collisions which are more likely at high pressures. Also, the Ar* metastables can play a significant role in generation of secondary electrons on the electrode surfaces \cite{Phelps2} and in discharge transient behavior \cite{Newton}. The metastables have been measured and used to establish kinetics of discharges for usually more complex dc geometries \cite{Kushner2005,Lazzaroni2010} or for rf plasmas \cite{Tochikubo1994,Tadokoro1998,Hebner1996,Rauf1997,Miyoshi2002}. The kinetics of metastables in other rare gases such as helium \cite{Wang2006,Prashanth2003} is also relevant. Measuring the number density and temporal evolution of these atoms is furthermore a crucial step in understanding energy transport mechanisms and in developing reliable models and precise simulations for these types of discharges. 

\section{Experimental setup}

\begin{figure}[tb]
\centering
\includegraphics[width=0.5\textwidth]{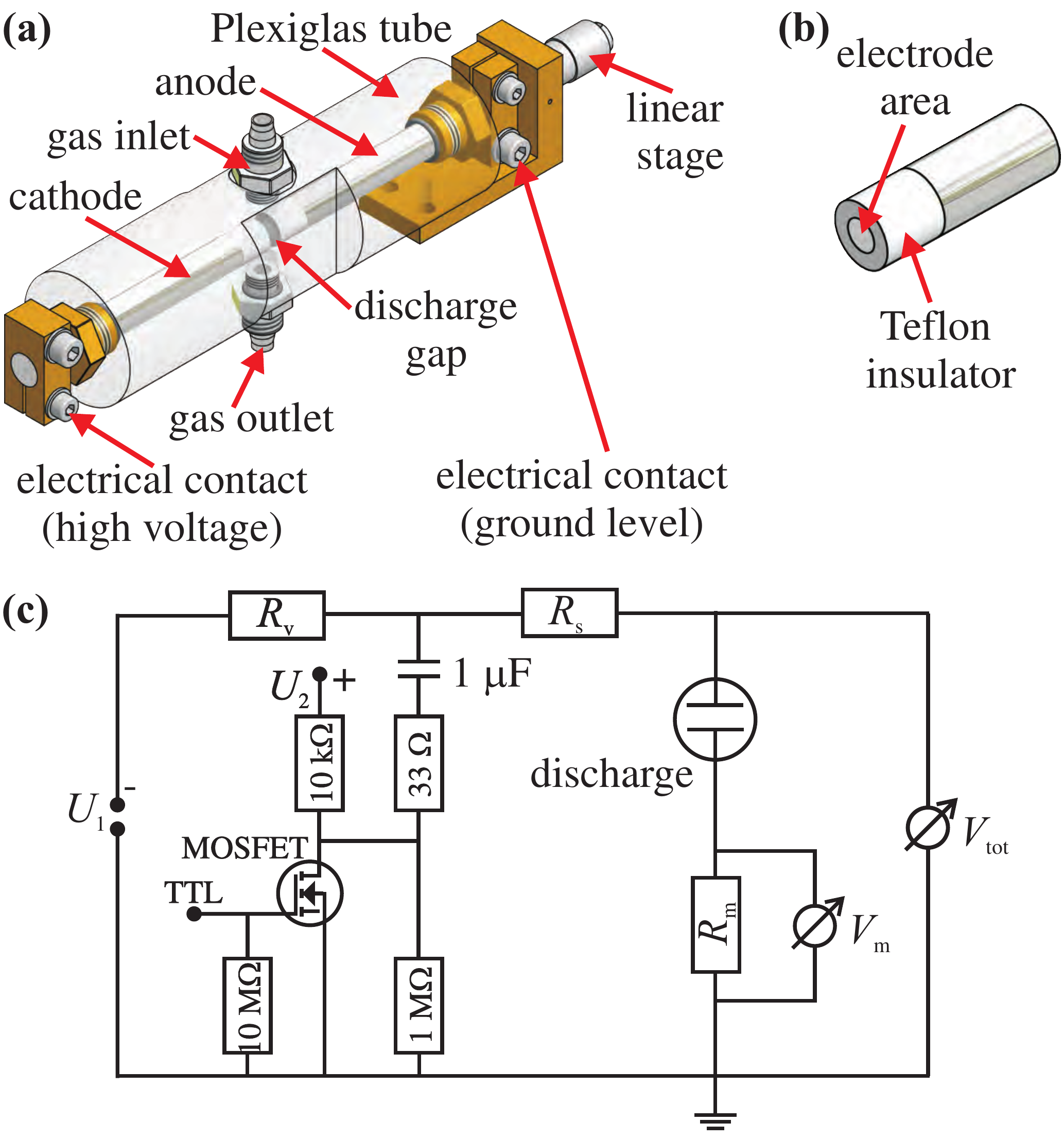}
\caption{(a) Schematics of the micro discharge chamber. The discharge gap is $d = 1\,\textrm{mm}$. (b) Sketch of one electrode end. The active electrode area has a diameter of $8\,\textrm{mm}$. (c) Sketch of the electrical circuit.}
\label{fig1}
\end{figure}

A sketch of the dc micro discharge chamber is shown in figure \ref{fig1}a. The plane parallel stainless steel electrodes are mounted within a tight fitting Plexiglas tube to avoid long-path breakdown \cite{Petrovic2}. The gas inlet and outlet are mounted at opposite sides of the discharge tube. A flux of Argon (typically $25\,$sccm) is used as feed gas to minimize the influence of impurities. The outer and inner walls of the Plexiglas tube are polished to improve the optical access to the plasma volume. The area between each electrode end and the Plexiglas tube is shrouded by a Teflon insulator as indicated in figure \ref{fig1}b. The discharge gap can be changed by a micro positioning linear stage, but was fixed at $d = 1\,\textrm{mm}$ during the experiments. Electrodes with a diameter of $8\,\textrm{mm}$ were used. The experiments were performed close to the Paschen minimum at $pd=1\,\textrm{Torr}\,\textrm{cm}$.

The electrical circuit is shown in figure \ref{fig1}c and is similar to the one presented in \cite{Maric2}. The voltage is monitored with a high voltage probe. The current is determined from the voltage drop over a monitoring resistor and corrected for the displacement current. To calculate the displacement current the capacity $C_{\textrm{\scriptsize{d}}}$ of the discharge is determined from measurements of the current and voltage during a short voltage pulse in vacuum. With the knowledge of $C_{\textrm{\scriptsize{d}}}$ the displacement current can be calculated:
\begin{eqnarray*}
I_{\textrm{\scriptsize{d}}} = C_{\textrm{\scriptsize{d}}}\frac{\textrm{d}U_{\textrm{\scriptsize{d}}}}{\textrm{d}t},
\end{eqnarray*}
where $U_{\textrm{\scriptsize{d}}}$ is the voltage drop over the discharge gap. 

Prior to each experiment the discharge is sustained at low current (roughly $10\,\mu\textrm{A}$) mode for around $15\,\textrm{minutes}$ until stable discharge conditions are ach\-ieved. During the experiments the discharge is first ignited in low current (a few $\mu\textrm{A}$) Townsend-like mode. Additionally, short voltage pulses (usually $<3\,\textrm{ms}$) are applied to change the discharge working point (intersection between a loading curve and micro discharge $V$-$A$ characteristics) to higher currents, as described in \cite{Jelenkovic}. Due to the short pulse length the discharge is running only for a short time in the high current mode, therefore significant gas heating and conditioning of the electrodes is avoided. 

\subsection{ICCD imaging}
\label{ICCD}

The axial light emission profile of the discharge is recorded with an ICCD camera (Andor, iStar DH734-18F-03). The camera objective (Rodenstock, Apo Rodagon D 2x) limits the observed wavelengths to the visible spectral range.

When images are recorded under steady state discharge conditions a sufficient gate width of the ICCD camera can be used to gather enough charge on the CCD chip. However, when monitoring the transient behavior the gate width of the camera has to be reduced down to the microsecond scale (in our experiments a gate width of $1\,\mu\textrm{s}$ was used). Unfortunately, under these conditions, a single shot records only a very noisy image because of the low light emission intensity. Therefore the charge gathered on the CCD chip is accumulated within a single pulse each time the discharge is running under identical conditions (the same current and voltage). An external delay generator is triggered on the signal of the discharge current each time the signal passes the falling slope of the set threshold voltage. The delay generator controls the gate position and gate width of the ICCD camera. 

\subsection{TDLAS setup}

\begin{figure}[tb]
\centering
\includegraphics[width=0.5\textwidth]{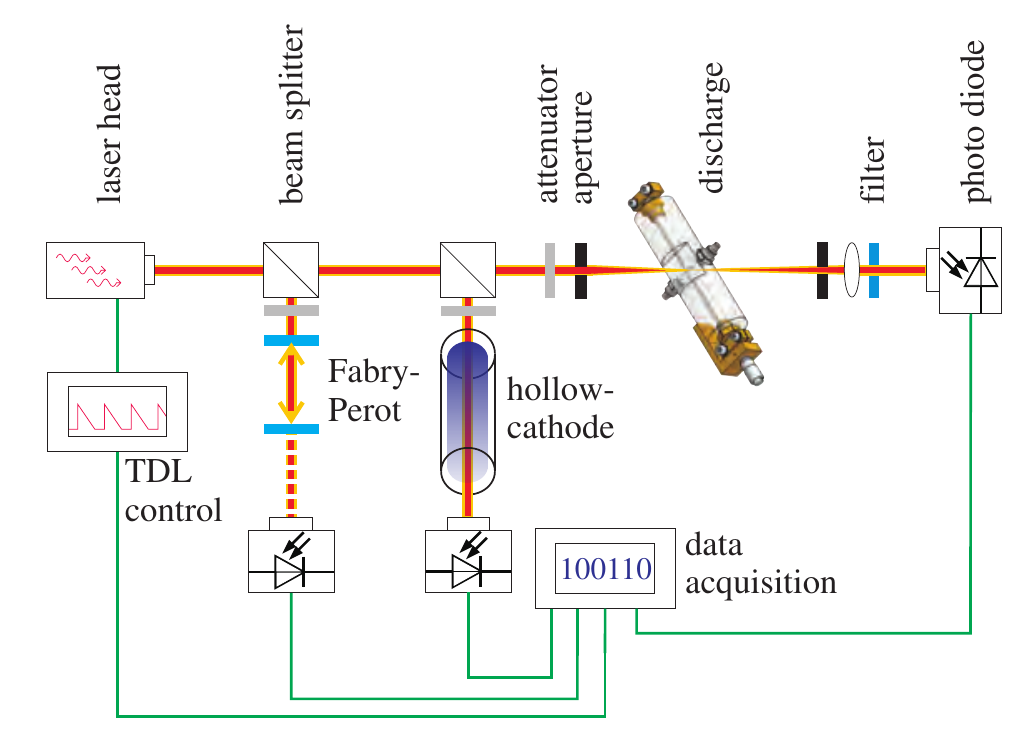}
\caption{Experimental Setup for TDLAS measurements.}
\label{fig2}
\end{figure}

The small dimensions of micro discharges and their operation at high pressures are a challenge for optical diagnostics, since high sensitivity and high spatial resolution are required. For the TDLAS measurements a diode laser with an external cavity in Littrow configuration is used. The linewidth of the laser ($<$10\,MHz) is much smaller than the width of the absorption line. Figure \ref{fig2} shows a schematic of the TDLAS experimental setup. The laser beam from the DL passes through two beam splitters. A part of the beam is guided to a Fabry-Perot interferometer (1\,GHz free spectral range) and another part through a low pressure reference cell, both necessary to calibrate the system. The part of the beam transmitted through the beam splitters into the discharge is attenuated by neutral density filters with an optical density in the order of 3, and guided into the discharge with a beam power of the order of a few $\mu$W, to avoid any saturation effects. After passing the discharge the beam is guided through a set of apertures and filters to suppress the emission from the plasma by reducing the collection angle and blocking wavelengths different than the observed transitions. The transmitted beam intensity is measured by a fast photodiode that provides a time resolution of a few hundred nanoseconds. The wavelength is tuned to the 1s$_5\rightarrow$\,2p$_9$ transition of Ar* at around 811.5\,nm. After recording the spectral profile of the absorption line, the metastable density $N_i$ is given by:
\begin{eqnarray*}
\int_0^\infty\ln\left [ \frac{I_0(\nu)}{I(\nu)}\right ]d\nu=S=\frac{\textrm{e}^2f_{ik}l}{4\epsilon_0 m_{\textrm{e}} c}\cdot  N_i,
\end{eqnarray*}
where $I(\nu)$ and $I_0(\nu)$ are the intensities of transmitted radiation with and without the presence of absorbing species, $l$ the path length through the absorbing medium, $S$ the area under the absorption curve that provides the line-averaged density of the absorbing species, $f_{ik}$ is the oscillator strength of the line and $N_i$ the density of the lower level \cite{Sadeghi}. All constants and oscillator strengths were taken from the NIST Atomic Spectra Database. Details about the calculation of metastable densities from the spectral profile are described in a previous publication \cite{niermann}. 

Due to the cylindrical geometry of the Plexiglas tube, the optical path through the discharge can be retraced only coarsely. For this reason the laser beam was by intention set up with poor spatial resolution to cover the complete active region of the discharge with a spot size of about $0.5\cdot 2$\,mm$^2$, providing us the discharge integrated metastable density. Taking the vertical divergence of the beam into account, the mean absorption length $l$ through the plasma was approximated to be 7\,mm.

\section{Results and discussion}

\subsection{Steady state volt-ampere characteristics and axial light emission}
\label{steady}

\begin{figure}[tb]
\centering
\includegraphics[width=0.5\textwidth]{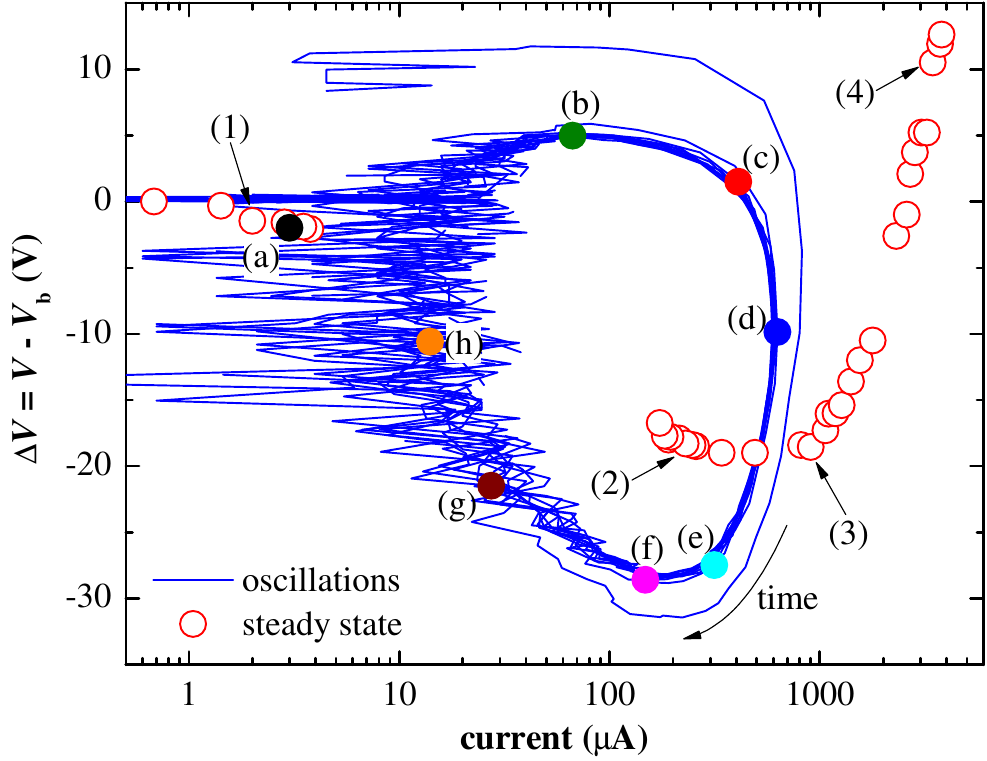}
\caption{Volt-ampere characteristics for steady state discharge (open circles) and during oscillations (solid line). Both experiments were performed at $pd=1\,\textrm{Torr}\,\textrm{cm}$. For better illustration the voltage is displayed as the difference between the discharge voltage ($V$) and the breakdown voltage ($V_{\textrm{\scriptsize{b}}}=220\,\textrm{V}$). Numbers (1)-(4) indicate the conditions of the 2D images for steady state discharge shown in figure \ref{fig4}. Solid dots marked with letters (a)-(h) refer to measurements during oscillations shown in figure \ref{fig5} and \ref{fig6}.}
\label{fig3}
\end{figure}

Figure \ref{fig3} shows a $V$-$A$ characteristics recorded under steady state discharge conditions. Between the low current Townsend-like discharge and the normal glow discharge a region of currents and voltages exists where no steady state regime can be reached for any discharge conditions. These instabilities were detected in standard size discharges and described through a combined effect of the external circuit and the effective negative differential resistance that may be observed in $V$-$A$ characteristics \cite{Petrovic1,Jelenkovic,Phelps,Stefanovic,Petrovic3}. The cause of negative differential resistance is the slightly increased electric field in front of the cathode. Under conditions (moderately high pressures) when electron production increases rapidly with $E/n$ (where $E$ is the electric field and $n$ the gas density) the small increase of the field in a narrow region in front of the cathode allows the field to be reduced below the uniform breakdown field in a much wider range leading to a slight decrease of overall voltage with an increase of current. The slightly increased field also affects the secondary electron yields and those become current dependent \cite{Phelps}.

In the region of instabilities the discharge can only operate in a transient regime, periodically switching from low current to high current mode \cite{Maric1}. This dynamic $V$-$A$ characteristics is indicated in figure \ref{fig3} by the blue line. The analysis of the axial emission and $V$-$A$ characteristic during the oscillations will be given in the next chapter.

2D images of the axial light emission profile at selected positions on the static $V$-$A$ characteristics (label 1-4) are presented in figure \ref{fig4}. The imaging optics was focused on the discharge axis of symmetry. As depth of field was only 0.13 mm, the obtained 2D images present the two dimensional light distribution of the emission form the plane that contains discharge axis and is perpendicular to the objective. The vertical axis represents the radial light distribution and horizontal axis the axial distribution. Additionally, the horizontal axis is several times enlarged compared to the vertical axis to improve the visibility. To precisely measure the discharge symmetry one would have to observe the radial light distribution through a transparent anode like in experiments in standard size discharges \cite{Maric1,Zivanov}. Nevertheless, some information on the radial emission from the recorded 2D images can be obtained from the graphs presented here.

In the low current diffuse Townsend-like mode (label 1) the discharge emission is exponentially growing from cathode to anode where it reaches its maximum. The discharge current is low enough that the space charge effect can not change the homogeneous external field significantly. The discharge spreads over the full electrode diameter and has the profile of a Bessel function. By the increase of discharge current the discharge changes to the normal glow (label 2). The peak of the axial distribution moves towards the cathode, indicating the formation of the cathode fall. In radial direction the discharge is highly constricted. The constrictions are shifted away from the center towards the electrode edge. The position of the constrictions is stable and is the result of the local variation of the secondary electron yield which seems not to be affected further by the discharge current. As current is further increasing (from 2 to 3) the current density stays constant and the discharge is spreading in radial direction. Between points (2) and (3) in figure \ref{fig3} the discharge is operating in the normal glow, similar to the low pressure, large scale discharges \cite{Maric1}. At point (3) the discharge occupies the whole electrode diameter, which marks the ending point of the normal glow. From this point the discharge runs in abnormal glow (label 4). The current increase leads to a higher light intensity.  The peak intensity shifts closer to the cathode and marks the cathode fall edge. All of these observations in our micro discharge are also typical for standard size discharges \cite{Maric1,Zivanov}. 

\subsection{Time resolved 2D emission distributions of discharge transients}
\label{freeruunning}

As noticed in the last chapter, there is a region of instabilities where the discharge is not stable. This oscillatory behavior is described through dynamic $V$-$A$ characteristics (figure \ref{fig3}, blue line). The characteristic hysteresis, previously reported for hollow cathode discharges \cite{Donko} and parallel plate standard size discharges \cite{Maric1,Maric3} for the relaxation oscillations is present: the discharge current starts from the low current regime and runs through the upper branch to the maximum current value and turns again to the low current. The time resolved voltage and current waveforms of transients are presented in figure \ref{fig5}. Both dynamic $V$-$A$ characteristics in figure \ref{fig3} and waveforms in figure \ref{fig5} show the same transient behavior of voltage and current except for the first transition from low to high currents. The first transition is different because the voltage first follows the shape as expected for operation under conditions of pristine gas without free charged particles, as seen in \cite{Maric1}. During later transitions the charged species and excited atoms remain from the previous discharge and thus reduce the breakdown voltage. Therefore the discharge switches to self sustained oscillations and the shape of the current and voltage signal are necessarily different. The periodic behavior allowed us to accumulate light signals from different transients within a single voltage pulse to record reliable images as described in section \ref{ICCD}. 

\begin{figure}[tb]
\centering
\includegraphics[width=0.5\textwidth]{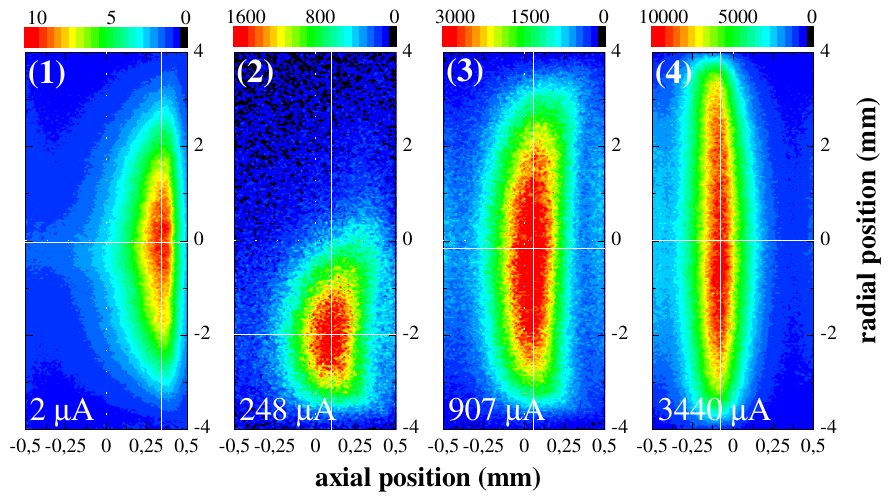}
\caption{2D images of the axial light emission profile of steady state discharge. Labels (1)-(4) correspond to the conditions indicated in figure \ref{fig3}. Cathode and anode are located at $-0.5\,$cm and $+0.5\,$cm respectively. Dotted lines mark the central axes of the discharge chamber, while solid lines mark the position of the peak of emission. The discharge current is shown in the bottom left corner of each image. The bar on top of each image indicates the discharge intensity recorded by the ICCD camera.\newline
(1) Townsend-like discharge\newline
(2)-(3) Normal glow discharge\newline
(4) Abnormal glow}
\label{fig4}
\end{figure}

\begin{figure}[tb]
\centering
\includegraphics[width=0.5\textwidth]{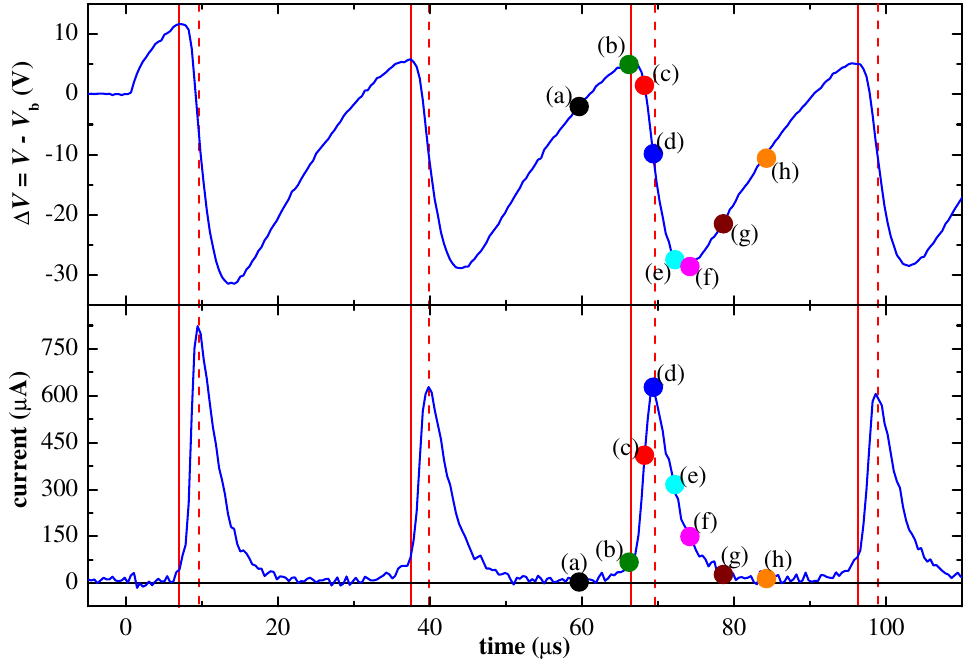}
\caption{Discharge voltage (with substracted breakdown voltage $V_{\textrm{\scriptsize{b}}}=224\,\textrm{V}$) and current as a function of time during oscillations ($pd=1\,\textrm{Torr}\,\textrm{cm}$). The corresponding $V$-$A$ characteristic is presented in figure \ref{fig3}. The solid dots (a)-(h) indicate the conditions of the 2D images shown in figure \ref{fig6} and the corresponding axial emission profiles at the peak of emission shown in figure \ref{fig7}. Dashed lines mark the positions of the maximum of each current peak while solid lines indicate the (positive) maximum of each voltage peak.}
\label{fig5}
\end{figure}

\begin{figure}[tb]
\centering
\includegraphics[width=0.5\textwidth]{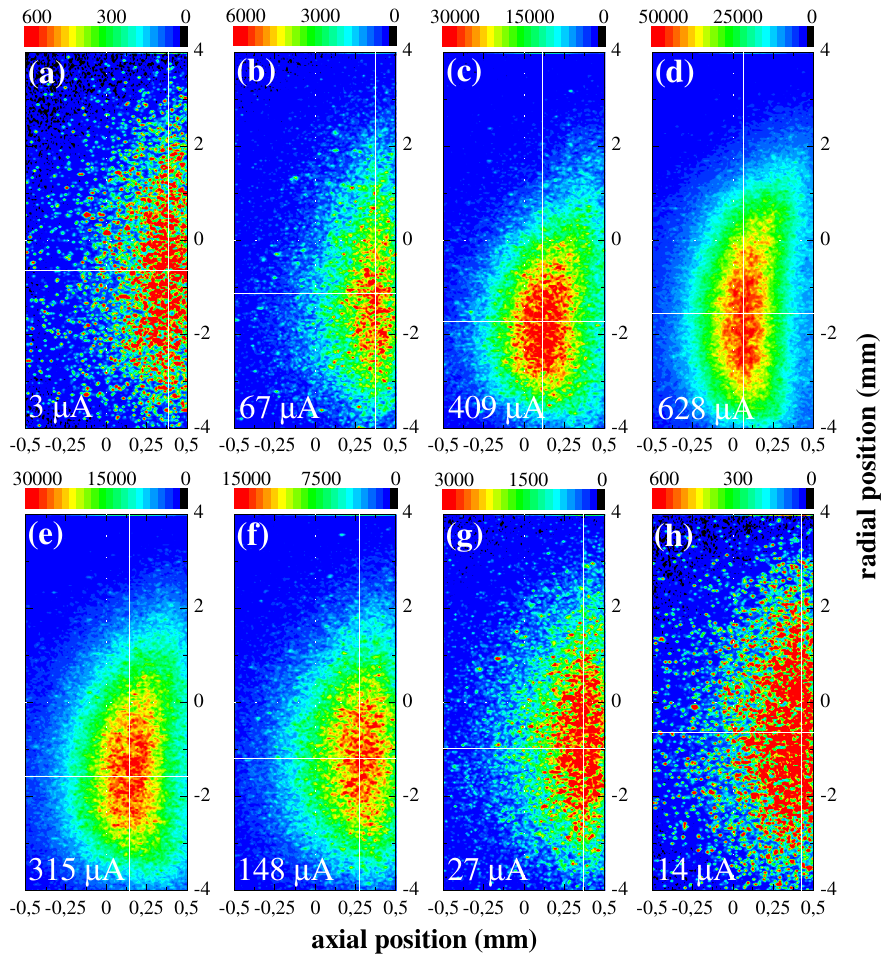}
\caption{2D images of the time development of the axial light emission during oscillations. Labels (a)-(h) correspond to the conditions indicated in figure \ref{fig5}. Cathode and anode are located at $-0.5\,$cm and $+0.5\,$cm respectively. Dotted lines mark the central axes of the discharge chamber, while solid lines mark the position of the peak of emission. The discharge current is shown in the bottom left corner of each image. The bar on top of each image indicates the discharge intensity recorded by the ICCD camera.}
\label{fig6}
\end{figure}

\begin{figure}[tb]
\centering
\includegraphics[width=0.5\textwidth]{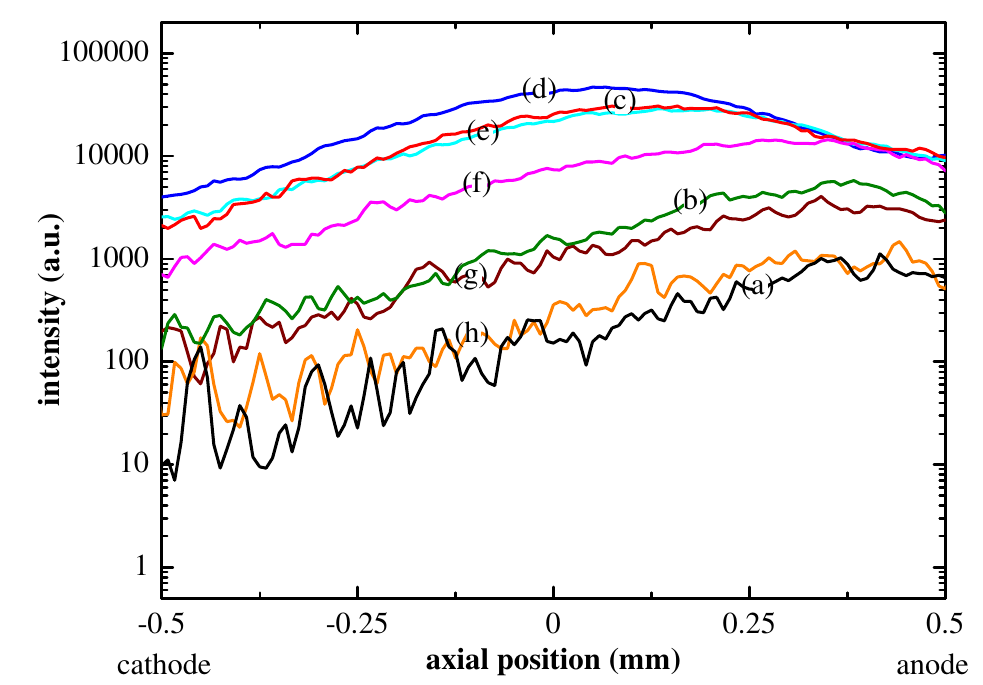}
\caption{Axial emission profiles at the peak of emission as indicated in figure \ref{fig6} for different images (a)-(h).}
\label{fig7}
\end{figure}

Figure \ref{fig6} shows 2D images of the axial light emission recorded by the ICCD camera at different discharge voltage and current values, as indicated by the labels (a)-(h) in figure \ref{fig5}. The corresponding axial light emission profiles at the peak of emission are presented in figure \ref{fig7}. In the transient regime the discharge develops from the low current Townsend-like diffuse mode to the high current normal glow mode. It has to be pointed out that images during the oscillations (figure \ref{fig6}) and in steady state (figure \ref{fig4}) have been recorded with different exposure times, therefore absolute intensities between these two measurements cannot be compared directly.

During transient Townsend-like mode (label a) the discharge is diffuse and the peak of emission is close to the anode. The discharge occupies the full electrode diameter. The light emission increases exponentially from the cathode to the anode, which is characteristic for the homogeneous electric field with negligible space charge effects. The current rises and space charge builds up slowly leading to a small drop of the voltage (label b). At the same time the light emission of the discharge increases and the peak of emission moves away from the anode. The cathode fall develops as the discharge current is rising further (label c). The peak of emission is shifted to the cathode and the discharge is highly constricted. Comparing the current values and the emission profile with the steady state conditions (figure \ref{fig4}b) we conclude that the discharge is operating in the normal glow. At the current maximum (label d) the light emission has reached its highest value and the peak of emission is located almost at the middle of the discharge gap. The discharge is less constricted in radial direction than in the previous state (label c), again characteristic of the normal glow. As current is dropping (label e and f), the peak of emission moves back from the center towards the anode, while the profile becomes more Bessel-like. Finally at low currents the discharge operates in the Townsend-like mode again (label g and h) as indicated by the diffuse discharge spread over the discharge diameter (figure \ref{fig6}) as well as the exponential increase of the light emission from the cathode to the anode (figure \ref{fig7}). Afterwards, this process repeats.

\subsection{Gas temperature under steady state discharge conditions}

The gas temperature and the metastable density have been determined from the Gaussian part of the line profile measured by TDLAS.

\begin{figure}[tb]
\centering
\includegraphics[width=0.5\textwidth]{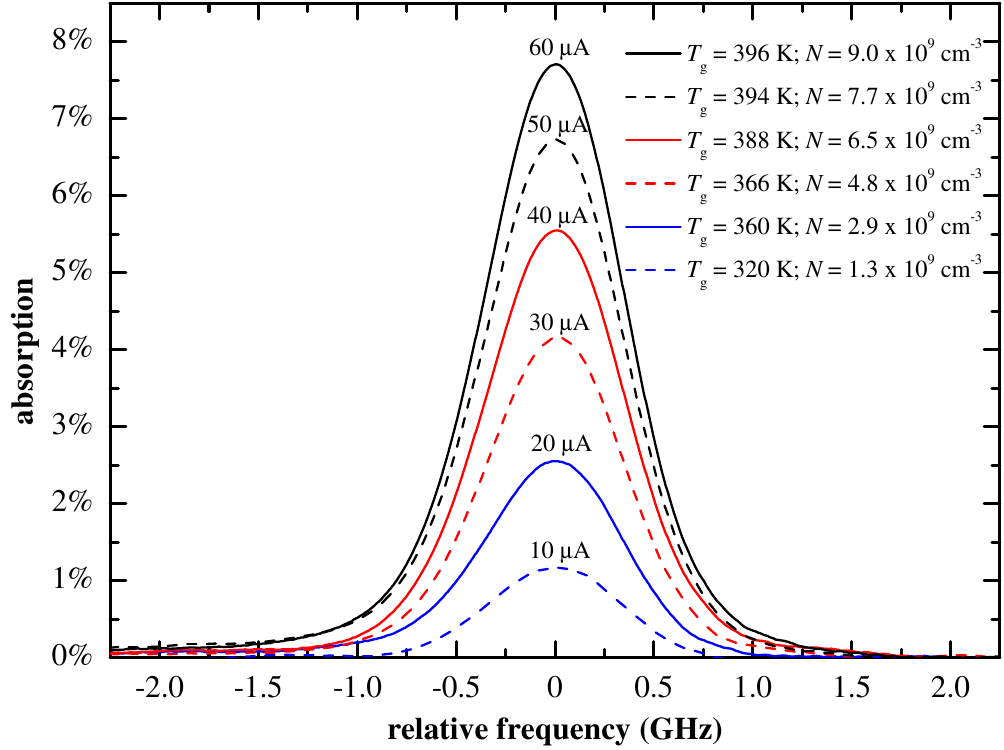}
\caption{Spectral profile of the Ar$^{\textrm{*}}$ 1s$_5\rightarrow$\,2p$_9$ metastable transition measured by TDLAS for different steady state low currents ($pd=1\,\textrm{Torr}\,\textrm{cm}$). The gas temperatures $T_{\textrm{\scriptsize{g}}}$ and the metastable densities $N$ have been determined from the Gaussian part of the line profile.}
\label{fig8}
\end{figure}

Figure \ref{fig8} shows examples of the absorption profiles of the Ar* 1s$_5\rightarrow$\,2p$_9$ metastable transition for varying discharge currents in the low current Townsend-like steady state mode. Measurements have been taken at $pd=1\,\textrm{Torr}\,\textrm{cm}$. 

The absorption line profiles are described by a Voigt-profile, a convolution of Gaussian and Lorentzian profiles, both having contributions in the same order of magnitude in this case. The area under the Voigt-profile is directly proportional to the absolute metastable density. The Lorentzian profile is primarily caused by pressure broadening and adds to the line profile with about 240\,MHz for a pressure of 10\,Torr. Stark broadening can be neglected in this case due to the low electron density. The profile, determined by the Doppler broadening, is strongly temperature dependent, and varies between 750 and 850\,MHz for typical discharge conditions. The Ar* 1s$_5\rightarrow$\,2p$_9$ transition is commonly used to measure the gas temperature from the Doppler width of the absorption line, since these metastable levels are in quasi-equilibrium with the ground state atoms, due to equipartition of kinetic energy between particles with similar masses in elastic and metastability exchange collisions \cite{Latrasse}.

The widths of the Gaussian component $\Delta\nu_D$, caused by Doppler broadening, is described by
\begin{eqnarray*}
\Delta\nu_D=\frac{2}{\lambda_0}\left[2\ln{(2)}\frac{k_{\textrm{\scriptsize{b}}} T}{M}\right]^{1/2},
\end{eqnarray*}
where $T$ is the temperature of the absorbing species (here assumed equal to the gas temperature), $M$ the mass of the species, and $\lambda_0$ the central wavelength of the observed transition \cite{Demtroeder}. For current spectra the temperature could be calculated with an uncertainty of less than 1\,\%.

Measurements reveal that with increasing discharge current the gas temperature rises from around ambient temperature up to $400$\, K. As electron density and temperature correlate with the discharge current, metastable densities go up as well, because the main excitation source for metastable species is provided by direct electron collisions.

Similar measurements have been performed in the high current steady state glow regime. The discharge current was varied between $186\,\mu$A (lowest possible current after oscillations) and $315\,\mu$A (maximum limit of the power supply). The gas temperature rises from 469\, K to 526\, K and the metastable density from $2.1\cdot 10^{10}\, \textrm{cm}^{-3}$ to $3.2\cdot 10^{10}\, \textrm{cm}^{-3}$.

\begin{figure}[tb]
\centering
\includegraphics[width=0.5\textwidth]{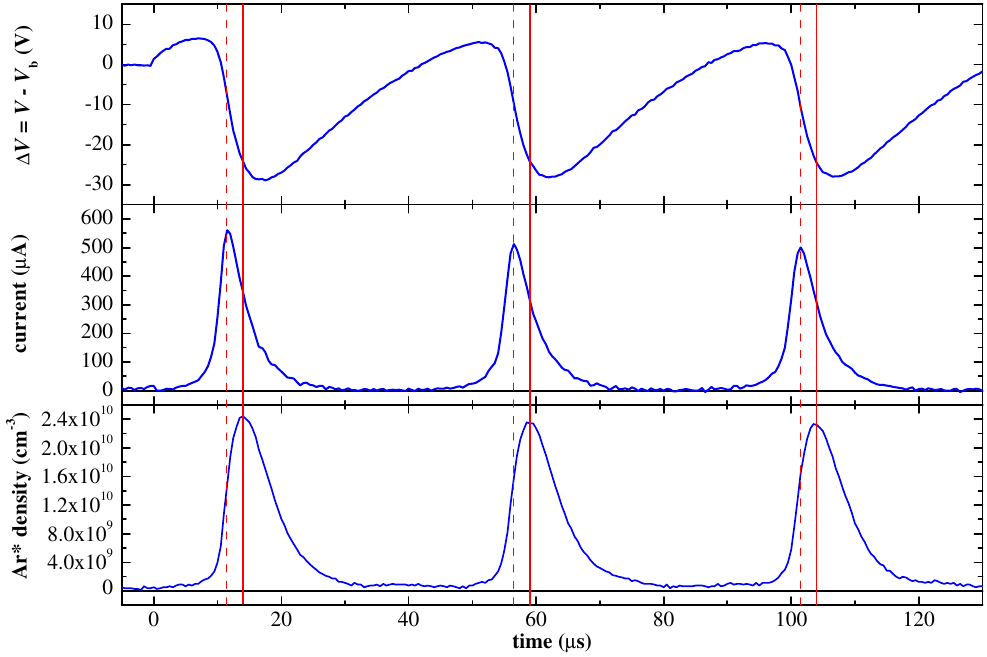}
\caption{Discharge voltage, current and argon 1s$_5$ metastable density as a function of time during oscillations ($V_{\textrm{\scriptsize{b}}}=229\,$V, $pd=1\,\textrm{Torr}\,\textrm{cm}$). Dashed lines indicate the maximum of each current peak while solid lines mark the maximum of the metastable density. The metastable lifetime is $\tau = 4.5\,\mu$s.}
\label{fig9}
\end{figure}

The dc results have somewhat greater but similar densities of the metastable atoms as the oscillating discharge transient (see figure \ref{fig9}). Thus the pulsed current peaking above 500$\,\mu$A yields a density of $2.4\cdot 10^{10}\,$cm$^{-3}$ while the dc glow discharge at 315$\,\mu$A leads to $3.2\cdot 10^{10}\,$cm$^{-3}$. As can be seen from the slope of the metastables density the majority of the metastables are produced at the peak of the current. The period when the current is above 50\% of the peak value (4$\,\mu$s) is less than 50\% of the duration of the metastables pulse. The effective current is thus around 270$\,\mu$A while the effective metastables density is around $1.3\cdot 10^{10}\,$cm$^{-3}$. Having in mind that the period of the metastables pulse is twice the duration of the current pulse the effective excitation rates in steady state and self-pulsed operation appear to be similar. Nevertheless it is evident that losses are high and thus in self-pulsed operation the losses would be twice as high as in dc operation which would require a more efficient production of metastables in the pulse. As the excitation coefficient increases rapidly with $E/n$ \cite{Bozin} and also instantaneous $E/n$ overshoots (see figure \ref{fig3}) the steady state values the results indicate an increased effective production of metastables in the oscillating mode as compared to the steady state conditions.

\subsection{Time development of metastable densities during oscillations}

The metastable densities presented in figure \ref{fig8}, that have been measured for low currents, show already that the species are a considerable source of potential energy in the system. The direct collisional electron excitation of Ar* from the ground state is a strong energy sink for electrons more energetic than the Ar* excitation threshold of 11.5\,eV.

The temporal metastable evolution, as shown in figure \ref{fig9}, is highly correlated with the rise of discharge current. The maximum of each current peak (indicated by dashed lines) coincides with the highest metastable production rate, which is given by the maximum of the first derivative. This is in agreement with the fact that the metastable atoms are produced mainly by direct electron impact excitation. Absolute densities reach maximum values  of about $2.4\cdot 10^{10}\, \textrm{cm}^{-3}$ (indicated by solid lines), $2.5\,\mu$s after the current peak, since the metastable excitation rate still exceeds the loss rate although the current is decreasing. The metastable decay during the decline of discharge current is a convolution of residual metastable production and the limited lifetime of the species. After the current ceases the decay is purely determined by the primary loss processes, namely diffusion, two- and three-body collisions with ground state atoms, and the Penning ionization loss due to impurities.

Metastable lifetime measurements were performed in the constant low current regime of the discharge, to exclude any influence of electrons on the de-excitation of the species. The lifetime values are given as the decay constant of an exponential function fitted to the density profile in the decaying tail. Under the given discharge conditions the metastable lifetime was measured to be 4.5\,$\mu$s.

Assuming infinite purity of the argon gas we can propose that the metastable lifetime is simply determined by diffusion to the walls and the two- and three-body collision processes with argon ground state atoms. For given discharge conditions metastable pooling can be neglected since their influence is about two orders of magnitude weaker. The dominant loss channels would therefore be:
\begin{eqnarray*}
\textrm{Ar}^* + \textrm{Ar} &\rightarrow& 2\textrm{Ar}\\
\textrm{Ar}^* + 2\textrm{Ar} &\rightarrow& \textrm{Ar}_2 + \textrm{Ar}\\
\textrm{Ar}^* + \textrm{wall} &\rightarrow& \textrm{Ar},
\end{eqnarray*}
leading to a calculated  metastable lifetime in pure argon ($p=10$\,Torr; $T=300$\,K) of:
\begin{eqnarray*}
\tau=(K_2\cdot N_{\textrm{\scriptsize{Ar}}}+K_3\cdot N_{\textrm{\scriptsize{Ar}}}^2+ D_{\textrm{\scriptsize{Ar}}}\cdot\Lambda^{-2})^{-1}, 
\end{eqnarray*}
with $K_2=2.3\cdot10^{-15}\,\textrm{cm}^3\textrm{s}^{-1}$  the rate coefficient for two-body collisions, $K_3=1.4\cdot10^{-32}\,\textrm{cm}^6\textrm{s}^{-1}$ the rate coefficient for three-body collisions, $N_{\textrm{\scriptsize{Ar}}}=3.3\cdot10^{17}\textrm{cm}^{-3}$ the argon ground state density at room temperature, $D_{\textrm{\scriptsize{Ar}}}=7.28$\,cm$^2$s$^{-1}$ the diffusion coefficient and $\Lambda=[(\pi/\textrm{Length})^2+(2.405/\textrm{Radius})^2]^{-1/2}$ the characteristic diffusion length in the discharge chamber \cite{Tachibana}. These coefficients produce rates at 10 Torr pressure of: 700 s$^{-1}$, 1500 s$^{-1}$ and 7500 s$^{-1}$ which leads to a calculated lifetime of $103\,\mu\textrm{s}$. The calculated lifetime is clearly larger by a factor of more than 20 than the measured value (4.5 $\mu\textrm{s}$). One should note that Molnar and Phelps found smaller values of these rate coefficients ($K_2=1.2\cdot10^{-15}\,\textrm{cm}^3\textrm{s}^{-1}$ and $K_3=0.85\cdot10^{-32}\,\textrm{cm}^6\textrm{s}^{-1}$) and that other but similar values are often used in the literature \cite{MolnarPhelps,Lymberopoulos,Gudmundsson,Gudmundsson2} but none of those values would suffice to put this lifetime based on the ground state atom quenching and diffusion in line with the experiment.  

As the measured lifetime values are much lower, the first instinct would be that  the discrepancy  can be attributed to the loss of metastable atoms by the excitation transfer  to impurities. Assuming that the dominant impurity contribution is due to residual N$_2$ molecules (the O$_2$ quenching rate coefficient is in the same order of magnitude), the impurity level can be estimated:
\begin{eqnarray*}
N_{\textrm{\scriptsize{N}}_2}=K_{\textrm{\scriptsize{N}}_2}^{-1}\cdot(\tau^{-1}-K_2\cdot N_{\textrm{\scriptsize{Ar}}}-K_3\cdot N_{\textrm{\scriptsize{Ar}}}^2-D_{\textrm{\scriptsize{Ar}}}\cdot\Lambda^{-2}),
\end{eqnarray*}
where $N_{\textrm{\scriptsize{N}}_2}$ is the molecular nitrogen density and $K_{\textrm{\scriptsize{N}}_2}=3\cdot10^{-11}\,\textrm{cm}^3\textrm{s}^{-1}$ the quenching rate coefficient of argon metastable atoms with nitrogen \cite{LeCalve}. Taking the measured lifetime into account, the impurity intrusion is in the order of 2\%. Taking the large surface to volume ratio and the low pressure of the discharge into account the estimated amount of impurities could be accepted but it is still excessive.

In addition to the impurities the large losses and therefore the short lifetime of the metastables may be explained by electron induced quenching, which has been established to be the main loss process in higher density plasmas, such as inductively coupled RF plasmas \cite{Masahiro}. Taking the total current and the effective area into account the electron density for an identical field distribution can be estimated as a function of $pz$ (where $z$ is the axial coordinate and $p$ is the pressure) by applying a hybrid calculation as used for the standard size discharge \cite{Maric4}. The estimated electron density is then in the order of $10^{12}\,\textrm{cm}^{-3}$. This may be coupled with a reasonable value of the electron induced quenching, which is in the order of 2-5$\cdot 10^{-7}\,$cm$^{3}$s$^{-1}$ \cite{Petrovic1995,Baranov1981,Bretagne1987}. Thus the equivalent lifetime is roughly $4.8\,\mu$s. This lifetime is consistent with the experimental observations. Electron induced quenching may proceed by collisional coupling of the metastables to the nearby radiative state (threshold less than $0.1\,$eV), collisional coupling to 2p states and ionization. It appears that electron induced quenching may be the dominant loss channel for metastables during the early afterglow in microdischarges even when diffusion is quite high due to larger surface to volume ratio.

Two further issues need to be resolved before making such a claim. The first is the maintenance of the electron density in the "afterglow" and the second is the thermalization of the electron energy below the threshold required to realize electron induced transitions to higher excited states which is in the order of 1-2$\,$eV. Modeling of breakdown delay times in argon reveals that the period when diffusion is ambipolar (and the losses of electrons are consequently relatively small) is relatively long and for our conditions it exceeds the measured lifetime by almost an order of magnitude \cite{Markovic2005}. On the other hand we have performed a Monte Carlo simulation of the thermalization of the electron energy distribution function EEDF in the afterglow \cite{Petrovic2007}. Starting from a typical EEDF with a mean energy of 4\, eV we have followed the time dependence of the electron population at different energies. The high energy tail decays very rapidly for less or around 1\,$\mu$s. However the decay to 2\, eV mean energy takes several microseconds and the decay to 100\, meV would take\, 0.5 ms. In other words during the period of decay of metastables there is a sufficient number density of electrons with a sufficient energy to maintain the electron induced quenching of metastables.

\section{Summary}

We have shown time resolved axial light 2D images of a parallel plate dc micro discharge in steady state as well as during discharge transients. The static $V$-$A$ characteristics is similar to the large scale, low pressure discharges, with distinguished low current diffuse mode, normal and abnormal glow. The measured axial distributions support this similarity between micro discharges and large scale, low pressure discharges. Between the low current mode and normal glow the region of oscillations has been found. During the relaxation oscillations the discharge develops from the low current mode (several $\mu$A) to the high current normal glow mode ($\approx 600\,\mu$A) repetitively. With increasing current the discharge intensity raises and the peak of emission moves away from the anode as the cathode fall develops. The normal glow has a constant current density and shows characteristic constriction of the conducting channel, which grows in diameter as the current is increasing.

The time development of the Ar* metastable densities in the discharge has been measured by tunable diode laser absorption spectroscopy. The discharge current and the metastable density are highly correlated. At the current maximum the highest metastable production rate can be observed. During the operation of the discharge electron induced excitation and eventually dissociative recombination produce a large population of metastables allowing stepwise processes that affect the ionization balance but are also the dominant metastables quenching channel. It is possible that impurities contribute but excessive abundance of N$_2$ of 2\% is required to explain the results. During the effective afterglow (the decaying part of the current and the period when the current is at a constant low value) electron induced quenching controls the rapid loss of metastables. Even though the high energy tail of the EEDF decays rapidly the mean energy decays slowly and it takes hundreds of microseconds to fall below the threshold for collisional coupling between metastables and resonant states.

In a recent paper published independently of this work a similar experiment was carried out with a more detailed model \cite{Belostotskiy2011}. The higher pressure of that work favored three body processes but in general they come to the same conclusions as we do, that the fast loss of metastables is dictated by the large electron induced quenching in the early phases of the afterglow.

Gas temperatures and Ar* metastable densities have been determined under steady state discharge conditions from the line broadening of recorded absorption profiles for the low current Townsend-like mode ($T_{\textrm{\scriptsize{g}}} = 320-400\,$K, $N = 1.3-9.0\cdot 10^{10}\, \textrm{cm}^{-3}$) and the high current glow mode ($T_{\textrm{\scriptsize{g}}} = 469-526\,$K, $N = 2.1-3.2\cdot 10^{10}\, \textrm{cm}^{-3}$).    

\ack 
This project is supported by DFG (German Research Foundation) within the framework of the Research Unit FOR1123, the Research Department 'Plasmas with Complex Interactions' at Ruhr-Universit\"at Bochum, DAAD Grant project 50430267 and Ministry of Science (MNTRS) of Republic of Serbia projects 171037 and 41011.


\newpage
\clearpage

\newpage
\clearpage

\newpage
\clearpage

\newpage
\clearpage

\newpage
\clearpage

\newpage
\clearpage

\newpage
\clearpage

\newpage
\clearpage

\newpage
\clearpage

\end{document}